\documentclass[journal=jpccck,manuscript=article,layout=twocolumn]{achemso}
\usepackage{graphicx} 
\usepackage{amsmath,amssymb}
\usepackage{xcolor}
\usepackage{xr}

\title{The Long and Winding Pore: Tuning Porosity in Templated Nanoporous Carbons}

\let\oldmaketitle\maketitle
\let\maketitle\relax

\author{Zacharie Waysenson}
\affiliation{Laboratoire de Physique de L'{\'E}cole Normale Sup{\'e}rieure, ENS-PSL Research University, CNRS, Sorbonne Universit{\'e}, Universit{\'e} Paris Cit{\'e}, F-75005 Paris, France}
\email{zacharie.waysenson@phys.ens.fr}

\author{Baptiste Barthe-{}-Gold}
\affiliation{Laboratoire de Physique de L'{\'E}cole Normale Sup{\'e}rieure, ENS-PSL Research University, CNRS, Sorbonne Universit{\'e}, Universit{\'e} Paris Cit{\'e}, F-75005 Paris, France}
\alsoaffiliation{Direction Générale de l’Armement, F-75015 Paris, France}
\email{baptiste.barthe--gold@phys.ens.fr}

\author{Arthur France-Lanord}
\affiliation{Sorbonne Universit{\'e}, Mus{\'e}um National d'Histoire Naturelle, CNRS UMR 7590, Institut de Min{\'e}ralogie, de Physique des Mat{\'e}riaux et de Cosmochimie (IMPMC), F-75005 Paris, France}

\author{Mathieu Salanne}
\affiliation[PHENIX]{Sorbonne Universit\'e, CNRS, Physico-Chimie des Electrolytes et Nanosystemes Interfaciaux, F-75005 Paris, France}
\alsoaffiliation[RS2E]
{R\'eseau sur le Stockage Electrochimique de l’Energie (RS2E), FR CNRS 3459, France}

\author{A. Marco Saitta}
\affiliation{Laboratoire de Physique de L'{\'E}cole Normale Sup{\'e}rieure, ENS-PSL Research University, CNRS, Sorbonne Universit{\'e}, Universit{\'e} Paris Cit{\'e}, F-75005 Paris, France}
\alsoaffiliation{Institut Universitaire de France (IUF), 75231 Paris, France}

\begin{document}
\twocolumn[
\begin{@twocolumnfalse}
\oldmaketitle

\begin{abstract}

Disordered materials pose a recurring challenge for molecular simulation: realistic structural models are often generated through thermal protocols in which several structural descriptors emerge simultaneously and cannot easily be varied independently. 
Nanoporous carbons constitute a particularly relevant example: pore size, wall order, curvature, and network tortuosity generally emerge as outcomes of the thermal preparation protocol, rather than being independently prescribed structural parameters.
Moreover, for practical applications, models of porous materials should ideally exhibit a \textit{percolating} pore network, capable of supporting end-to-end transport of adsorbed species through the material.
To this end, here we introduce a hierarchy of templating strategies, built around a common melt–quench–anneal protocol driven by a machine-learned interatomic potential for carbon, that progressively separates these variables. A rigid cylindrical template interacting with the environment through a tunable Lennard-Jones potential provides independent control over pore size and local wall organization, while leaving the structural disorder of the surrounding carbon matrix essentially unchanged. Generalizing the template to a spring-connected, tortuous backbone further allows pore geometry and curvature to be prescribed without qualitatively altering the local templating mechanism established for the straight pore. Finally, making the template mobile lets the growing carbon network reshape it in turn, which sharpens pore boundaries and allows sub-nanometer pore sizes to be targeted precisely, as needed for applications such as supercapacitor electrodes. 
Ultimately, this approach enables systematic studies of structure–property relationships in porous materials for applications such as energy storage, desalination and gas capture.

\end{abstract}
\vspace{1em}
\end{@twocolumnfalse}
]

\section{Introduction}

Disordered nanoporous carbons are central to modern electrochemical and separation technologies. Their high specific surface area, electronic conductivity and chemical stability make them the electrode material of choice for electrochemical double-layer capacitors (supercapacitors), high-power storage devices that complement batteries in grid buffering, regenerative braking and pulsed-power applications \cite{Simon_2020,pohlmann2022a}. The discovery that specific capacitance increases when pore sizes fall below the solvated ion diameter \cite{Chmiola_2006} sparked two decades of research into confinement-controlled charge storage, leading to hierarchical and sub-nanometer porous carbons designed to accelerate ion transport and increase energy density~\cite{du2025a}. The same tunable porosity is exploited in capacitive deionization and blue-energy harvesting~\cite{simoncelli2018a}, as well as in ion separation~\cite{dai2026a}, CO$_2$ capture~\cite{mapstone2025a}, and in gas separation through carbon molecular sieves~\cite{wang2026critical}, mostly in pressure swing adsorption nitrogen generators. Across these applications, performance depends not only on the amount of porosity but on its geometric organization: pore size, wall curvature, surface disorder and network tortuosity act as distinct design levers whose combined effects on transport are only beginning to be understood.

Isolating these effects in simulations relies on the capability of accessing physically realistic carbon models whose structural descriptors can be controlled independently. 
Systematic studies of liquids in porous carbons have been made possible by recently developed tools\cite{Dick_2023}; however, the geometries and dynamics of the carbon matrix are usually idealized.
Techniques aimed at reproducing experimental observables have emerged~\cite{Peng_2026}, yet melt--quench and annealing molecular dynamics remains the standard approach: using classical bond-order potentials \cite{Palmer_2010,Thompson_2017,Sarkar_2024,detomas_2018,monti_2021} or, more recently, machine-learned interatomic potentials (MLIPs) with near-quantum accuracy \cite{Wang_2022,Qamar_2023}, it reproduces the pair correlation functions, ring statistics and hybridization ratios of real carbide-derived and activated carbons. However, the pore network then emerges from the cooling trajectory rather than being imposed. Changing the pore size through quench parameters therefore also alters local disorder, defect populations and wall curvature \cite{Yin_2022}, which precludes controlled numerical experiments in which a single variable is changed. This limitation is critical because experiments indicate that structural disorder is itself a key driver of capacitance \cite{Liu_2024}, that pore network tortuosity, rather than average pore size or total porosity, governs charging rates \cite{Kress_2026}, and that the vibrational modes of the electrode framework affect in-pore electrolyte organization and charging kinetics \cite{Waysenson_2025}. Disentangling these contributions requires a synthesis or simulation route that pure quench/anneal protocols cannot offer \cite{Gogotsi_2003,Forse_2015}.

Templating provides a natural way to decouple pore geometry from the rest of the carbon structure. Experimentally, \emph{hard templating} grows or infiltrates a carbon precursor around a rigid scaffold, typically a zeolite or silica, which is then etched away to leave a pore network that faithfully inherits its geometry \cite{Braun_2018,chi_2024,Aumond_2022}. Computationally, the scaffold can be replaced by excluded-volume regions inserted into a melt--quench trajectory and later removed, so that pore size is set by the interaction length scale rather than by the cooling protocol \cite{Luo_2021,Zhou_2025,Ngambia_2024}. More recent approaches control pore morphology through an implicit continuum field, such as a Gaussian random field or a Cahn--Hilliard phase field, subsequently relaxed with an MLIP, which allows porosity and pore-size distribution to be controlled at fixed density \cite{Yu_2025,Wang_2026a}. 
\emph{Soft templating}, in contrast, relies on a mobile, self-assembling structure-directing agent, usually a block copolymer or surfactant, that co-assembles with the precursor before being removed; it is the preferred experimental route to hierarchical, interconnected mesoporous networks \cite{ENTERRIA201679}. Its computational counterpart remains immature: the closest study couples surfactant self-assembly with the condensation of a silica network \cite{Jorge_2007}, rather than embedding a mobile template in a melt--quench simulation of amorphous carbon.

None of these approaches treats tortuosity as an independent structural parameter. Hard templates fix it through their chosen geometry, quench/anneal protocols make it an uncontrolled outcome of the cooling kinetics, and implicit-field methods provide no straightforward, physically transparent way to tune it at fixed pore size and chemistry \cite{Wang_2024,Li_2023,Wang_2026b}. Yet tortuosity is now recognized as a first-order determinant of rate capability, on par with pore size itself.

This gap motivates the methodology developed here. 
In a recent study\cite{Waysenson_2026}, we showed that the pore-size distribution of disordered graphitic nanoporous carbons can be empirically tuned, and used this novel technique to quantitatively relate pore size to capacitance. Here, we formalize the underlying strategy and generalize it beyond pore size to a broader set of structural descriptors.
Starting from a quench/anneal benchmark, we first introduce a rigid cylindrical template with a tunable Lennard-Jones interaction to control pore size and wall order independently. We then generalize it to a chain of particles connected by springs that imposes a controlled tortuosity on the resulting pore, and finally allow the template to become soft and mobile during the quench. More generally, this hierarchy offers a route to realistic atomic-scale disordered materials in which selected structural variables are controlled independently ,rather than emerging together from the preparation protocol.

\section{Quenching carbons}
\label{sec:quench}

Every structure generated in this work is produced with the same underlying melt-quench-anneal protocol, using the molecular dynamics software LAMMPS~\cite{Thompson_2022} in conjunction with a published Atomic Cluster Expansion (ACE) MLIP parameterization for carbon~\cite{Qamar_2023}. 

An initial configuration with carbon particles, placed randomly with a minimal distance criterion of 1.5 \AA, is first melted and equilibrated in the canonical ensemble~\cite{martyna1992a} at $T=6000$~K for $100$~ps, well above the melting point, so that all memory of the initial lattice is removed (Figure~\ref{fig:hard-quench-grid}). The system is then cooled at a constant rate of $10~\text{K}\,\text{ps}^{-1}$. For the reference (``no hold'') protocol, this ramp proceeds directly to $300$~K. 

We noticed that such a simple protocol causes numerous defects to be frozen in the carbon, especially sp chains. We therefore added an an annealing phase during the quench, with the aim of reducing the amount of defects and inducing a higher graphitization degree. This is achieved by pausing the ramp at an intermediate hold temperature $T_{\text{hold}}$ for $100$~ps before cooling resumes towards $300$~K, where the structure is further equilibrated. In practice we settled for $T_{\text{hold}}=3900~\text{K}$, in the temperature range where fragments begin condensing into curved graphitic
sheets.

\begin{figure}[ht]
    \centering
    \includegraphics[width=0.9\linewidth]{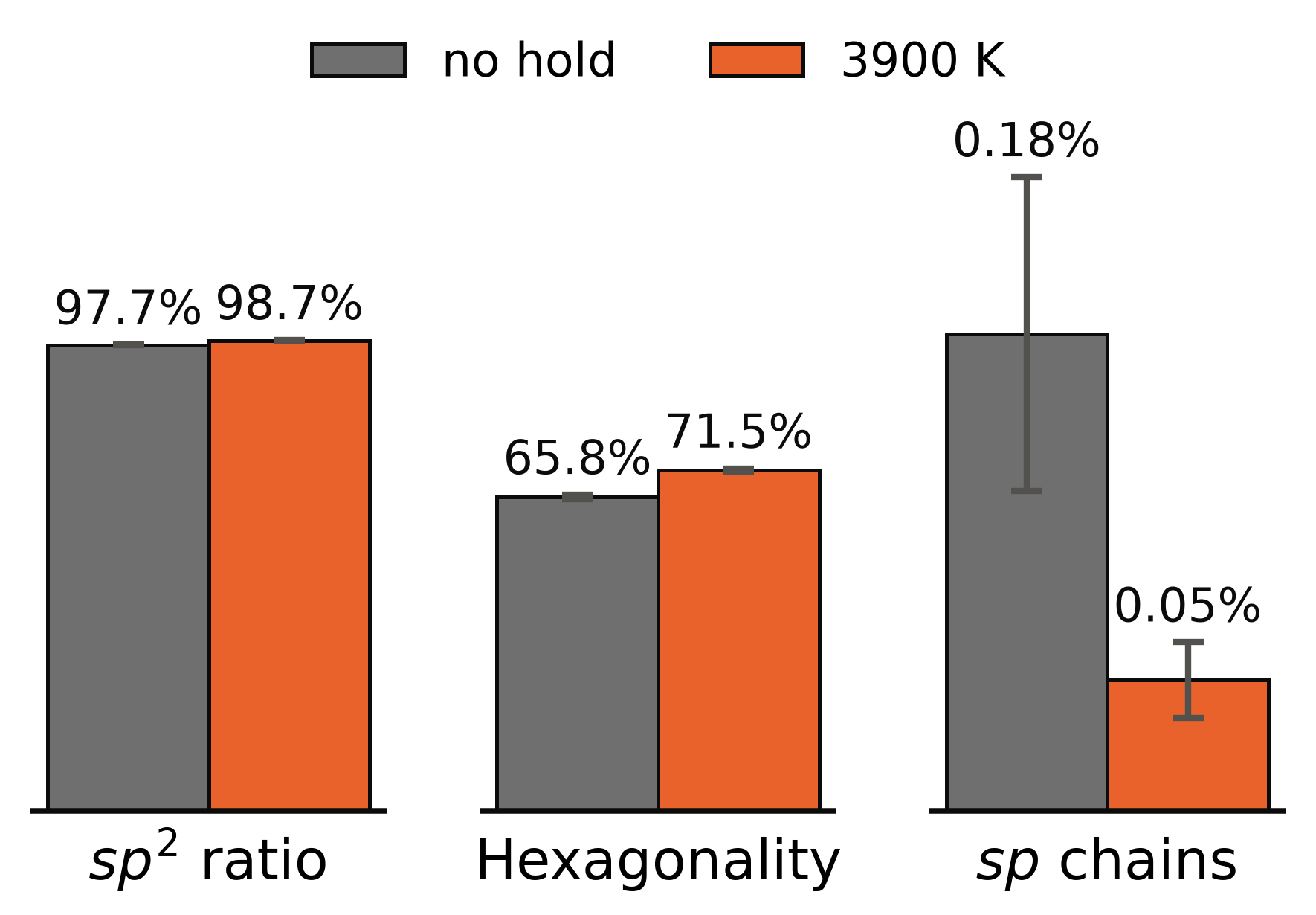}
    \caption{Structural features of quenched carbons with and without an annealing hold at 3900K. From left to right: proportion of sp$^2$ coordinated carbons, proportion of hexagonal rings, proportion of atoms involved in a long sp chain ($\ge3$ atoms). The metrics are averaged across 5 realizations.}
    \label{fig:quench-structure}
\end{figure}

Figure~\ref{fig:quench-structure} reports three structural properties of the generated carbons: 
the fraction of three-coordinated (sp$^2$) carbon atoms, the
fraction of six-membered (hexagonal) rings, and the sp chain proportion, defined as the proportion of atoms involved in a sp chain (a chain of 2-coordinated atoms) of length at least 3. These cover bridges between structures, dangling chains and floating closed loops..
Introducing the hold plateau at $3900~\text{K}$ improves the graphitization of the quenched structure (sp$^2$ and hexagonal rings counts increase) while reducing defects (see also Table S1), the C–C pair correlation function is essentially insensitive to the hold (Figure S1).

We therefore chose to use the procedure with the hold plateau as a general thermal profile for the rest of this work.

Although physically sound, disordered quenched structures do not guarantee percolation of the pore network required for practical applications. The following sections introduce a template into the quench protocol to ensure percolation through the structure, and to gain control over several aspects of the porous network.

\section{Hard templating}
\label{sec:hard}

\begin{figure*}[ht]
    \centering
    \includegraphics[width=0.8\linewidth]{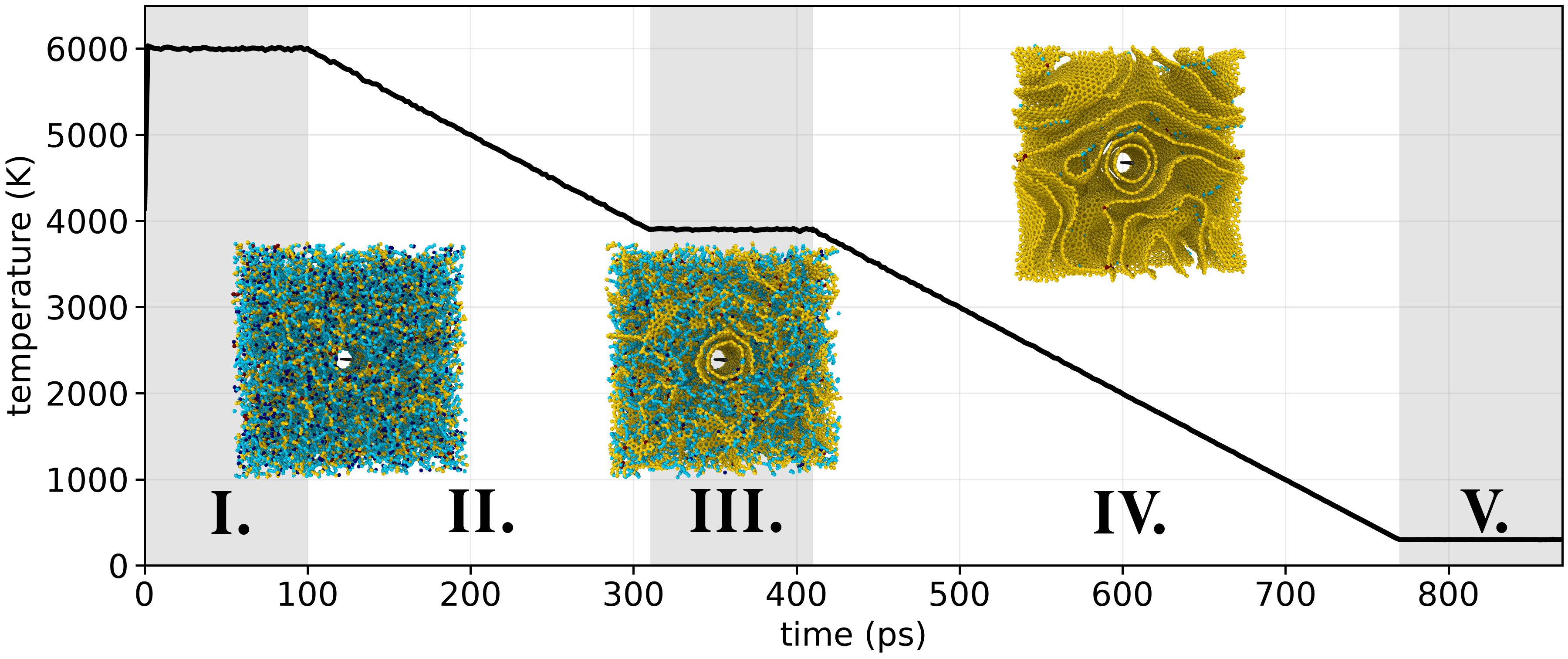}
    \includegraphics[width=0.8\linewidth]{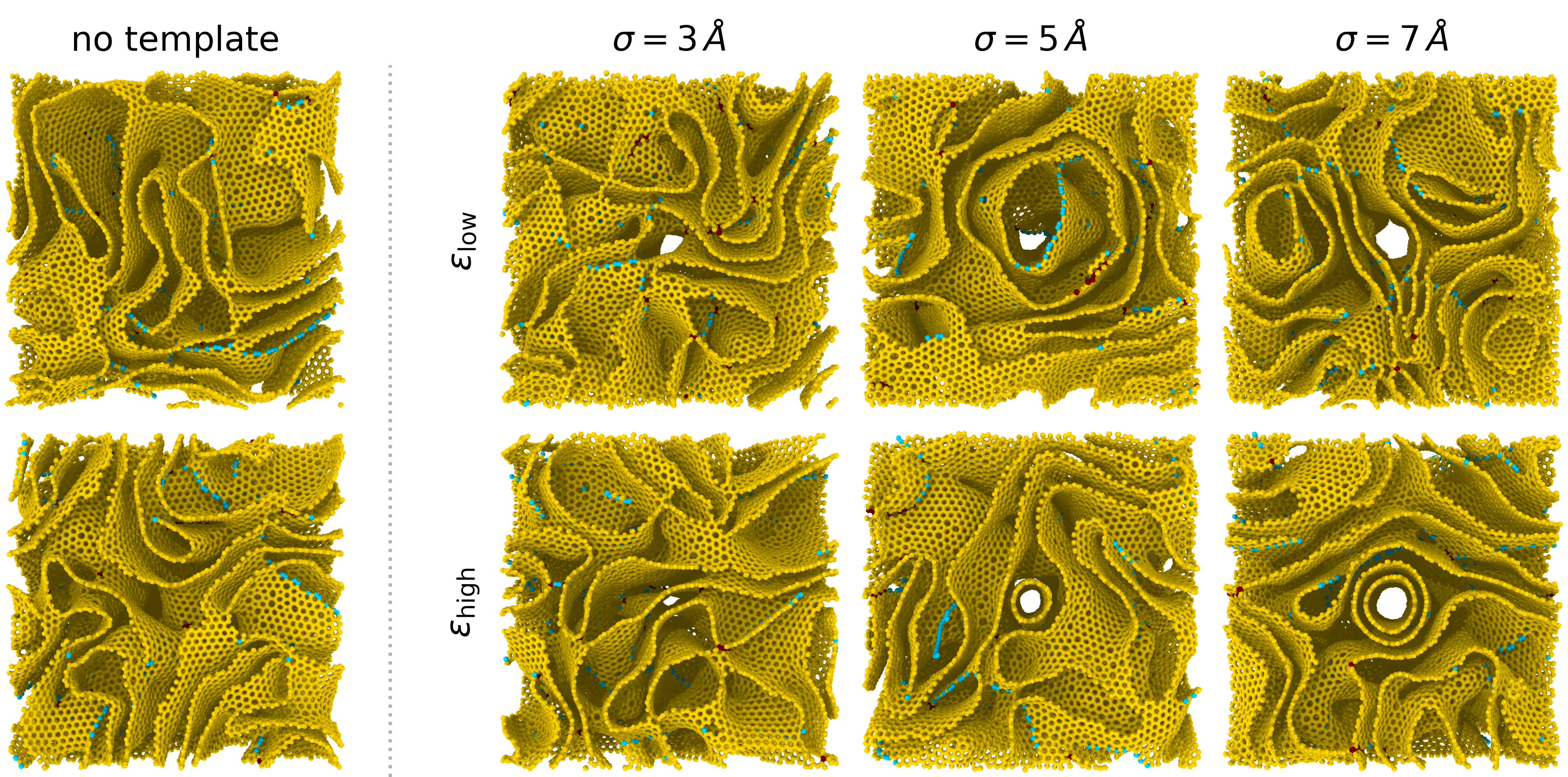}
    \caption{Generation of templated disordered carbons.  
(Top) Quench temperature protocol: I. Equilibrate liquid carbon around the template at 6000~K for 100~ps. II. Quench linearly at 10~K~ps$^{-1}$ to 3900~K. III. Hold at 3900~K for 100~ps. IV. Quench linearly to 300~K. V. Remove the template and equilibrate at 300~K for 100~ps.  
(Bottom) Structures obtained for different LJ parameters. The two leftmost are bulk reference structures (no template). Colors indicate carbon coordination: deep blue (0–1 bond), light blue (sp), yellow (sp$^2$), red (sp$^3$); template atoms are white.}
    \label{fig:hard-quench-grid}
\end{figure*}

We first introduce a simple model template: a linear, rigid chain composed of fictitious particles, spanning the $z$-axis of the simulation box, present throughout the same protocol described above and removed only when reaching $300~\text{K}$ before the final carbon-only equilibration at ambient temperature (Figure \ref{fig:hard-quench-grid}). 
Carbon atoms interact with the template (tpt) through a Lennard-Jones (LJ) potential,
\[
  U_{\text{C-tpt}}(r) = 4\varepsilon\left[\left(\frac{\sigma}{r}\right)^{12} - \left(\frac{\sigma}{r}\right)^{6}\right],
\]
so that the interaction range $\sigma$ and well depth $\varepsilon$ can act as two independent
control parameters, in contrast to a literal atomistic scaffold (e.g.\ a zeolite or silica lattice) whose
pore geometry is fixed by its crystal structure \cite{Braun_2018,chi_2024,Aumond_2022}.

Figure~\ref{fig:hard-quench-grid} displays quenched carbons for $\sigma\in\{3,5,7\}$~\AA\ at two well depths, $\varepsilon_\text{low}=0.01\,k_BT_\text{hold}$ and $\varepsilon_\text{high}=0.1\,k_BT_\text{hold}$, alongside the bare, non-templated reference. 

The two parameters control qualitatively different aspects of the resulting pore. Increasing $\sigma$ at fixed $\varepsilon$ simply increases the excluded volume:  the straight pore percolating through the \textit{z}-axis enlarges continuously as $\sigma$ is increased. The well depth $\varepsilon$, for its part, governs the organization of the carbon walls in the surroundings of the excluded volume. At $\varepsilon_\text{low}$, the template acts as a purely repulsive excluded-volume body, as a consequence the pore boundary remains
ragged and merges irregularly into the surrounding carbons. At $\varepsilon_\text{high}$, the same $\sigma$ values instead produce a clean, circular pore wrapped in one or more concentric graphitic shells. In this regime, the deep potential well acts as a nucleation seed that organizes carbon atoms into curved, nanotube-like walls on top of excluding a defined volume.

\begin{figure*}[ht]
    \centering
    \includegraphics[width=0.9\linewidth]{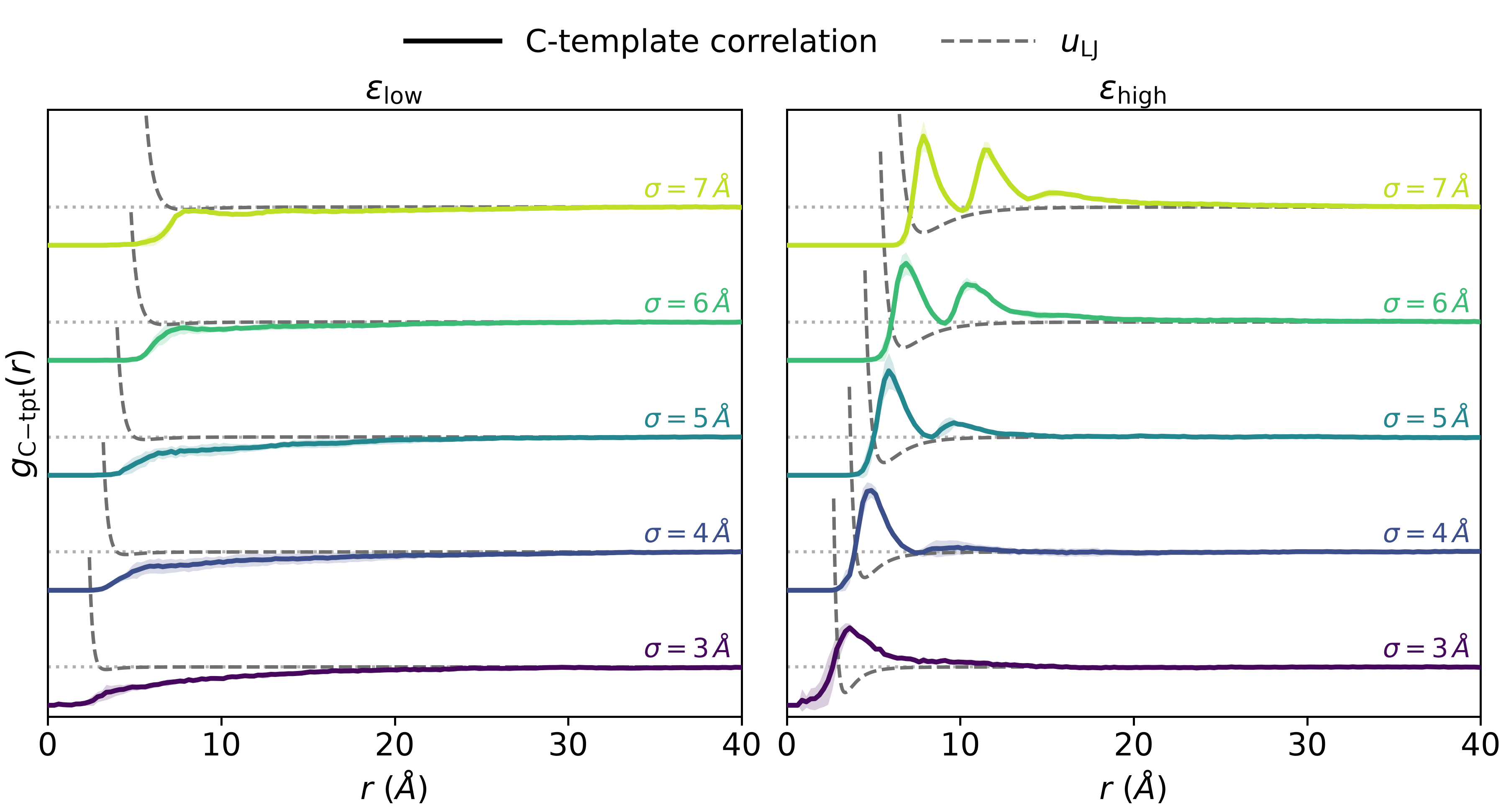}
    \caption{Carbon-template cross-correlation for the straight templates, at different values of the LJ potential, averaged over 5 realizations. LJ potential as reference (dashed lines).}
    \label{fig:hard-gr-template}
\end{figure*}

A quantitative characterization of the regimes shown in Figure~\ref{fig:hard-quench-grid} is provided by the carbon–template cross-correlation, $g_\text{C-tpt}(r)$, presented in Figure~\ref{fig:hard-gr-template} alongside the LJ potential as a reference.

At $\varepsilon_\text{low}$, $g_\text{C-tpt}(r)$ rises smoothly and monotonically from zero at the repulsion boundary set by $\sigma$ up to the bulk density value of 1: beyond the excluded volume, carbon atoms simply pack up without any regular layering, consistent with the disordered pore walls observed above. At $\varepsilon_\text{high}$, by contrast, a pronounced first adsorption peak develops just outside the exclusion radius, while a second and a third weaker peaks emerge as $\sigma$ increases. These peaks are clear signatures of one, and then several concentric carbon shells forming around the template near the LJ potential minimum, 
i.e.\ the nanotube-like walls seen directly in Figure~\ref{fig:hard-quench-grid}. 
In both regimes, the entire cross-correlation curve translates to larger $r$ as $\sigma$ increases, confirming that pore size (the exclusion zone) is controlled by $\sigma$ while the wall organization is controlled by $\varepsilon$. 
The emergence of a second or third nanotube layer as $\sigma$ increases at fixed $\varepsilon_\text{high}$ occurs because the LJ well width depends on both $\sigma$ and $\varepsilon$. Thus, increasing $\sigma$ broadens the potential minimum and allows additional layers to nucleate.

The introduction of the template also has a mechanical consequence that should be kept in mind. Since the simulations are performed at fixed cell volume, the excluded volume generated by the template reduces the configurational volume available to the carbon atoms and thereby introduces an additional local confinement, particularly near the carbon–template interface. Such coupling between geometrical constraints and the spatial distribution of molecular-scale stress is a general phenomenon: in earlier first-principles studies of linear and knotted polyethylene under tensile loading, geometrical constraint was found to strongly redistribute strain energy and determine where rupture ultimately occurred.\cite{Saitta1999Nature,Saitta1999JCP,Saitta2001JPCB}.
In the present case, as shown in Figure S4, we quantify the stress induced by the templating agent at $\varepsilon_\text{high}$. We show that the template removal allows the systems to relax, except for a residual stress in the $z$ direction.

Indeed, a rigid, straight cylinder is an idealized geometric model of a pore: it allows for connectivity and diameter control, but lacks branching or bending, which are two physically important features of realistic porous networks. 
The next section improves on this concern by turning the cylinder into a tortuous template.

\section{Tortuous hard templating}
\label{sec:tortuous}

To move beyond the perfectly straight pores, we  now allow the particles of the template chain to interact with each other in order to introduce bending and therefore tortuosity. The LJ interaction between the template and the carbons stays as is, while we add intra-template interactions: a two-body, harmonic spring between each bonded neighbors, $U_{\text{bond}}(r) = \frac{1}{2}k_{\text{spring}}(r-r_0)^2$ with a resting length of $r_0=2$~\AA, fixing the size of each link of the chain, and a three-body, angular potential acting on each triplet of bounded atoms,
$U_{\text{angle}}(\theta) = \frac{1}{2}k_{\text{angle}}(\theta-\theta_0)^2$. The resting angle $\theta_0$ prescribes the stiffness of the chain, or how curved its resting position will be. As $\theta_0$ approaches $180^\circ$, the chain becomes progressively taut and adopts a helical conformation. Conversely, as $\theta_0$ decreases, the chain exhibits increasingly localized bending, ultimately yielding a highly tortuous, spike-like morphology. In the present study, we therefore set $\theta_0 = 170^\circ$ throughout, as this value produces a regular S-shaped configuration consistent with tortuous pore geometries.
Finally, in order to avoid the chain collapsing onto itself, a repulsive term is added between pairs of unbounded atoms, $U_{\text{rep}}(r) = \frac{1}{2}k_{\text{rep}}(r-r_c)^2$ for $r<r_c$, with $r=3$~\AA.
The different harmonic constants were chosen as\\
$    k_\text{spring}=5\,k_BT_\text{hold}~\text{\AA}^{-2}\\
     k_\text{angle}=3\,k_BT_\text{hold}~\text{rad}^{-2}\\
     k_\text{spring}=6\,k_BT_\text{hold}~\text{\AA}^{-2}
$

At fixed bond resting length, one can tune the total length of the template chain by varying the number of its particles. 
As an example, in a simulation box of length $L_z =100$~\AA\ and template bond length $r_0=2$~\AA, using 50 particles results in a stiff chain of length $L_\text{template}$ stretched along the box. The tortuosity of the template can be defined as $\tau=\frac{L_\text{template}}{L_{z}}$, with $\tau=1$ for 50 particles. Doubling the number of template particles will result in a 200~\AA\ long chain with a corresponding $\tau=2$. Note that changing tortuosity also changes the template volume in the simulation, since more tortuous chains need more particles.
At a fixed simulation-cell volume, increased tortuosity mechanically enhances the excluded volume and, consequently, increases the degree of mechanical confinement imposed on the carbon phase.

Upon setting these parameters, a template-only equilibration protocol is executed prior to the introduction of the carbon atoms. This step of 100~ps NVT dynamics at 3900~K is intended to allow unconstrained diffusion and conformational relaxation of the template, resulting in a transition from an initially unstable straight-rod configuration to a bent, tortuous chain conformation.
At this stage, the template will remain fixed during the quenching procedures; the effects associated with allowing unrestricted interactions with the carbon atoms are addressed in the following section.

\begin{figure*}[ht]
    \centering
    \includegraphics[width=0.9\linewidth]{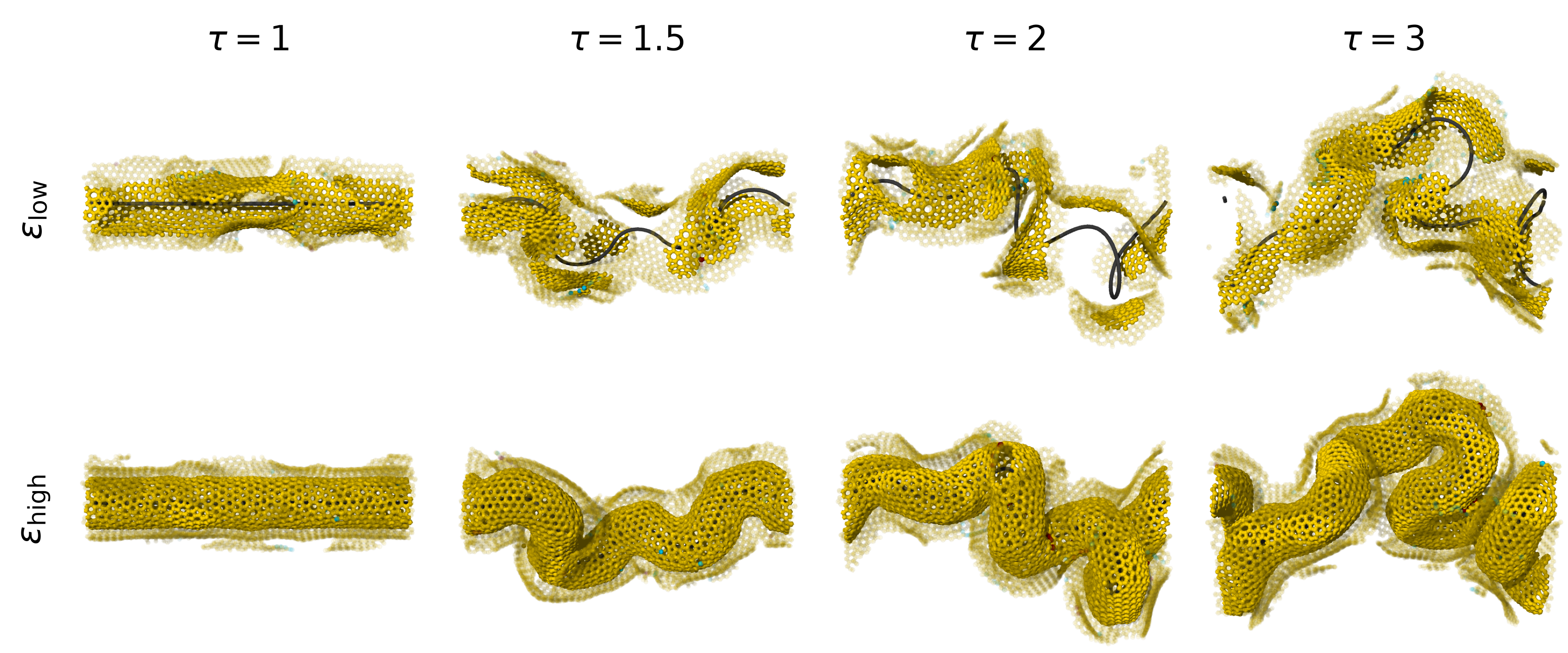}
    \includegraphics[width=0.9\linewidth]{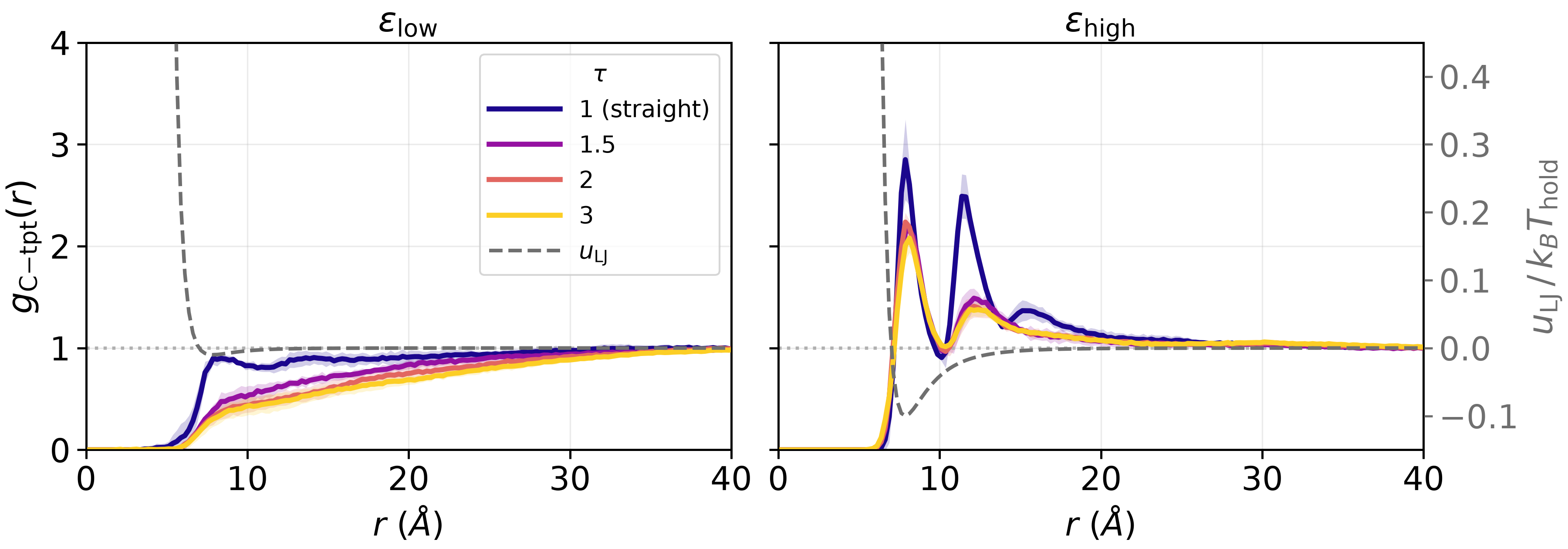}
    \caption{Templates of different tortuosities at $\sigma=7$~\AA, $\varepsilon_\text{low}$ and $\varepsilon_\text{high}$. 
    (Top) Visualization of the main pore formation around the template. Tortuosity 1 corresponds to the straight template of the previous section. The template atoms are colored in dark grey. The carbons within the first $g_\text{C-tpt}(r)$ peak ($r<10$~\AA) are opaque, those within the second peak ($10-15$~\AA) are transparent and the ones further than 15~\AA\ have not been rendered.
    (Bottom) Carbon-template cross-correlations at different tortuosities (full lines), LJ potential as reference (dashed lines).}
    \label{fig:tort-render}
\end{figure*}

\begin{figure*}[ht]
    \centering
    \includegraphics[width=0.9\linewidth]{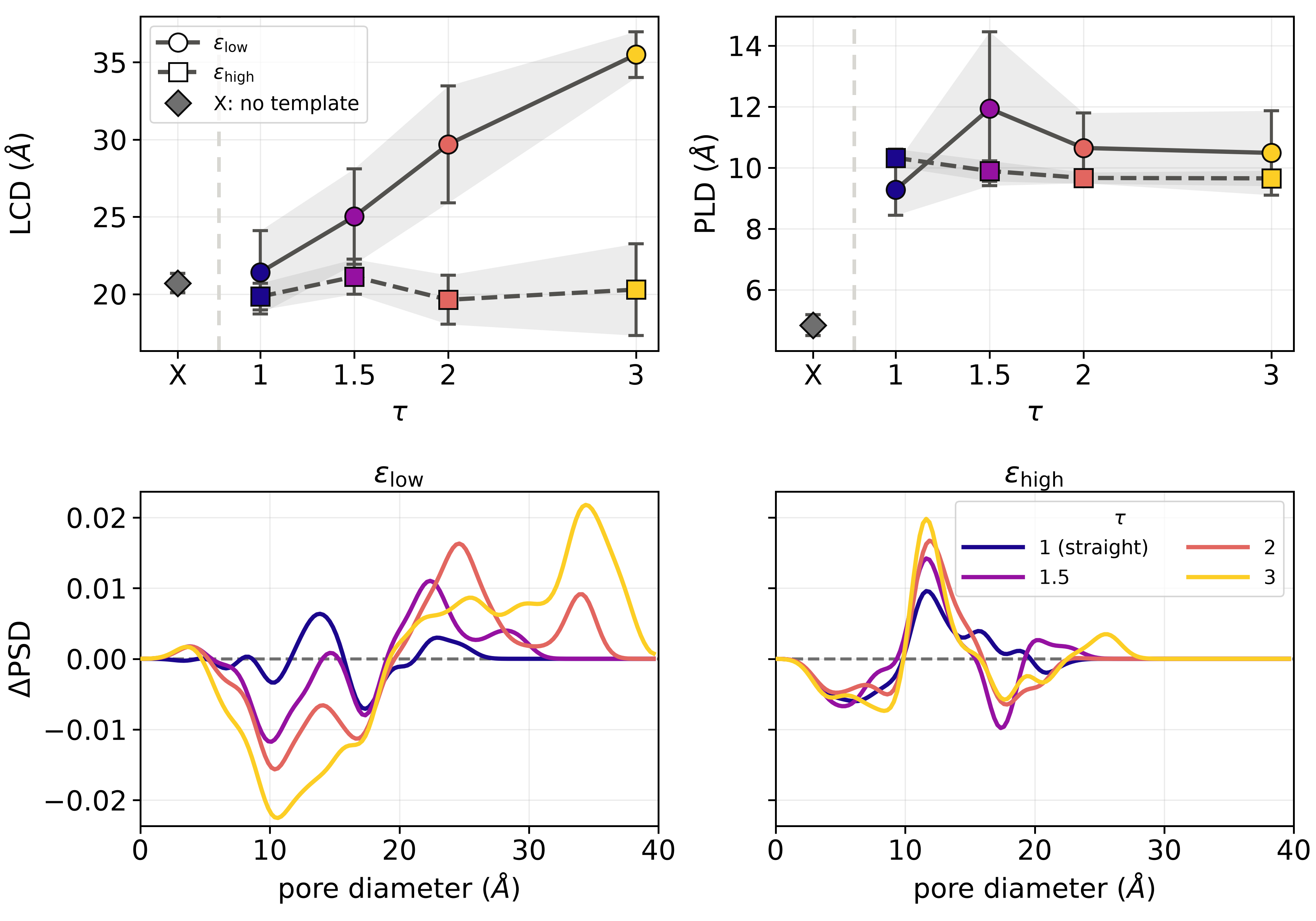}
    \caption{
    (Top) Largest Cavity Diameter (LCD) and Pore Limiting Diameter (PLD) of the carbons at different tortuosities for $\sigma=7$~\AA.
    (Bottom) Residual of the Pore Size Distribution (PSD) at different tortuosities ($\Delta\text{PSD}=\text{PSD}-\text{PSD}_\text{ref}$) smoothed with a 1~\AA\ gaussian kernel.}
    \label{fig:tort-psd}
\end{figure*}

Figure~\ref{fig:tort-render} (top panel) shows carbon organization around templates with tortuosities from 1 (straight case) to 3. The template–atom LJ parameters are fixed at $\sigma=7$~\AA\ with $\varepsilon_\text{high}$ and $\varepsilon_\text{low}$, and for each tortuosity, the same template is used for both $\varepsilon$ values. The method therefore controls pore bending during the quench: the straight $z$-axis tube (tortuosity 1) becomes S-shaped at 1.5 and can form loops at higher tortuosity. These pores retain the straight-case features: connectivity and percolation along $z$, an exclusion zone set by $\sigma$, and wall organization governed by $\varepsilon$ while the local carbon structure is unchanged (Figure S3, left).

The bottom-panel of Figure~\ref{fig:tort-render} displays a comparison of the carbon–template cross-correlation for different tortuosities at $\varepsilon_\text{high}$ and $\varepsilon_\text{low}$, with $\sigma=7$~\AA.
At $\varepsilon_\text{low}$, the behavior remains similar to the straight template: the interaction is essentially repulsive, creating an exclusion zone with cross-correlation 0, followed by a rise toward the bulk value 1 beyond the LJ minimum. With increasing tortuosity, however, the distance needed to recover to 1 becomes much larger. Figure~\ref{fig:tort-psd} provides an explanation \textit{via} the Largest Cavity Diameter (LCD): higher tortuosity yields larger pores because the repulsion excludes carbon from the template region without constraining it to a fixed nanotube radius. Carbon instead forms disordered graphitic sheets while avoiding the template, and in highly tortuous cases the chain can coil and enlarge the exclusion region (e.g. $\tau=2$ at $\varepsilon_\text{low}$ in Figure~\ref{fig:tort-render}). These larger pores lead to the higher degree of cross-correlation.\\

At $\varepsilon_\text{high}$, the cross-correlation indicates that nanotube-like ordering is preserved along the template bends: the first peak remains well defined, while the second peak is weaker and broader, suggesting tortuosity disrupts regular organization of the second layer compared with the straight case.
Consistently, the LCD in Figure~\ref{fig:tort-psd} stays roughly constant with tortuosity, since the nanotube layers inhibit the opening of larger pores, highlighting the contrast between weakly attractive (small $\varepsilon$) and strongly attractive (large $\varepsilon$) regimes.

Figure~\ref{fig:tort-psd} also displays the Pore Limiting Diameter (PLD), i.e., the diameter of the largest sphere that can percolate through the sample. Adding a $\sigma=7$~\AA\ template strongly increases the PLD relative to the bulk case, from $<6$~\AA\ to $>10$~\AA, which is the goal of the protocol. The PLD varies little with $\varepsilon$ or tortuosity, unlike the LCD. 
This suggests that $\sigma$ mainly controls the percolating diameter of the final carbon structure, largely independently of other parameters, even when larger pores open.

Finally, the $\Delta\text{PSD}$ in Figure~\ref{fig:tort-psd} provides more detailed information on the pore-size distribution. The pore-size distribution (PSD) is defined here as the probability density of pore diameters within the sample. We define $\Delta\text{PSD}$ as the difference between the PSD obtained in the presence of a template and the corresponding bulk PSD in the absence of templating. Consequently, a positive $\Delta\text{PSD}$ at a given diameter indicates that pores of that diameter occupy a larger volume fraction than in the bulk reference. Because both PSDs are normalized to integrate to unity, a positive feature in $\Delta\text{PSD}$ must be compensated by a negative feature at other diameters.

This representation further highlights the qualitative differences between
 the $\varepsilon_\text{low}$ and $\varepsilon_\text{high}$ regimes. 
At $\varepsilon_\text{low}$, increasing tortuosity produces a pronounced enhancement in the large-diameter tail of $\Delta\text{PSD}$, particularly for pore sizes in the range $20$--$40$~\AA. This trend is consistent with the LCD analysis: under a purely repulsive interaction, tortuous templates promote the formation of larger pores at the expense of intermediate pore sizes ($5$--$20$~\AA).

By contrast, at $\varepsilon_\text{high}$ the $\Delta\text{PSD}$ exhibits a markedly different response. Rather than an enrichment of large pores, the dominant change with increasing tortuosity is the growth of a peak centered around $10$--$15$~\AA. This feature corresponds to the percolating pore generated by the template, whose diameter is narrowly constrained by the concentric carbon shells. The modest increase of this peak with tortuosity can be attributed to the larger template volume fraction present in more tortuous systems, such that a greater fraction of the pore network is directly shaped by the templating process.

Overall, $\Delta\text{PSD}$ reveals an additional distinction between the two $\varepsilon$ regimes. For purely repulsive interactions, templating increases the prevalence of larger pores in a comparatively uncontrolled manner, even while maintaining control over the percolating diameter (PLD). At higher $\varepsilon$, both the PLD and the broader PSD remain constrained; the PSD develops a well-defined peak governed by the attractive well and the carbon shells formed within it.

Having established a detailed understanding of the behavior of multiple parameters in the rigid (hard-frozen) template, the following section examines the changes that occur when the template is allowed to evolve freely and interact with carbon species during quenching, rather than being constrained.

\section{Soft templating}
\label{sec:soft}

\begin{figure*}[ht]
    \centering
    \includegraphics[width=0.9\linewidth]{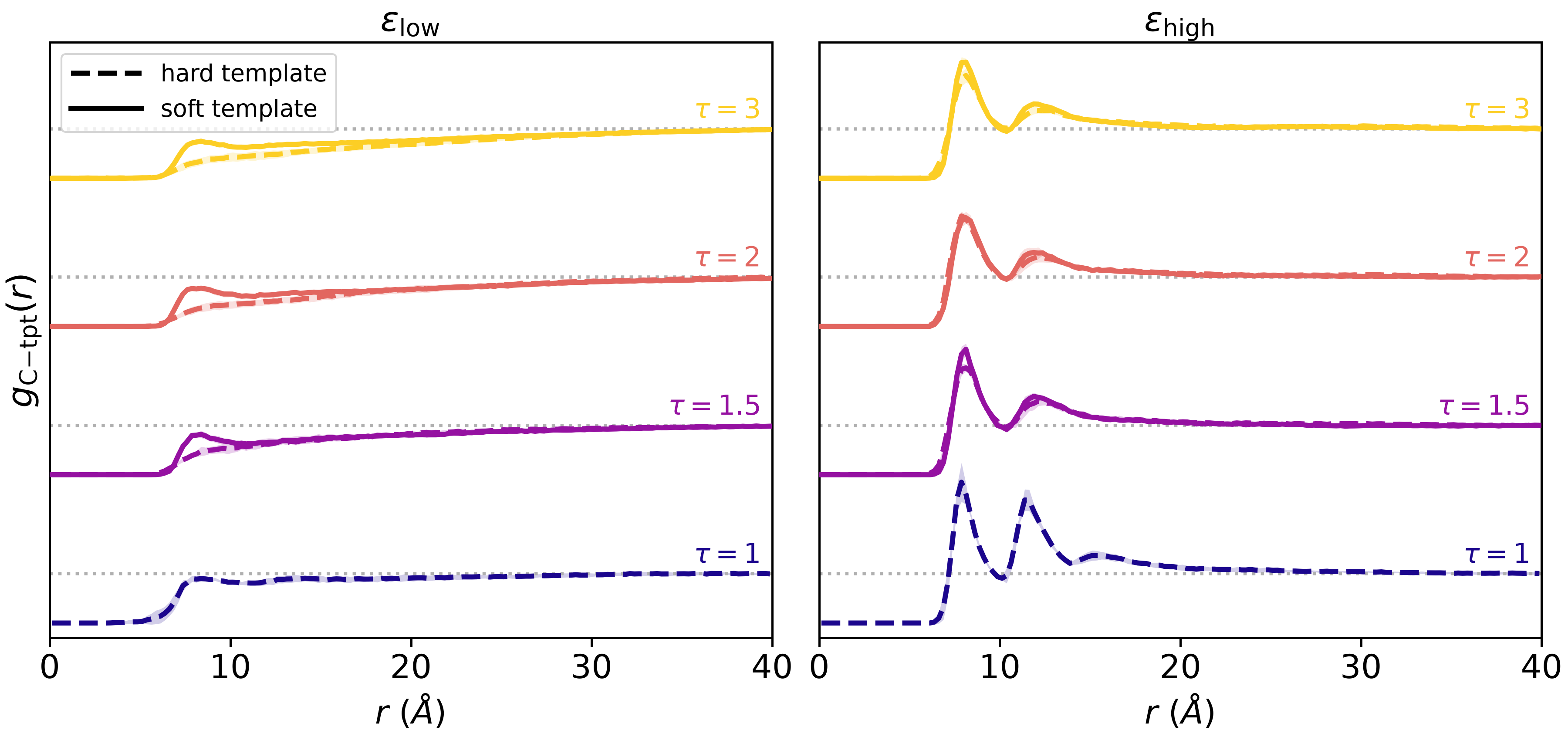}
    \caption{Carbon-template cross-correlation comparison for soft (mobile) and hard (frozen) templates at $\varepsilon_\text{low}$ and $\varepsilon_\text{high}$ with $\sigma=7$~\AA\ and different tortuosities.}
    \label{fig:soft_metrics}
\end{figure*}

In previous sections, the template particles were held fixed throughout the quench, thus imposing the geometry of the carbon matrix. We now go a step further and present a protocol in which the template particles can move along the quench. The intra-template interactions are unchanged, and the LJ potential now acts on both the carbon atoms and template particles. This lets the graphitizing carbon network locally rearrange the template, so the coupled carbon–template system can relax toward a more favorable configuration.
Figure~\ref{fig:soft_metrics} compares the carbon–template cross-correlation obtained with soft and hard templates.
For $\varepsilon_\text{low}$, the effect is pronounced: with the soft template, the cross-correlation rises more sharply just beyond the exclusion zone, indicating a higher carbon density directly adjacent to the repulsive region of the potential. This confirms that, with a hard template, the voids around the chain are widened by the organization of graphitic sheets around an immobile chain in addition to the LJ repulsion itself. When the chain is mobile, it can adapt to the newly formed graphitic sheets, which increases the cross-correlation after the exclusion zone. Hence, with a soft chain, not only does the template impose a tortuous and percolating path in the carbon, but the reciprocal action of carbon on the template also sharpens the boundaries of the pore.

For $\varepsilon_\text{high}$, the soft template yields a slightly higher first adsorption peak and a narrower second peak, although this effect is subtle compared with the previous regime. This is explained, with the hard template, by the already well-defined first carbon shell around the chain. Granting the template mobility in this case does not greatly impact the dual carbon-template organization, because carbon is already well structured within the allowed volume.

In both regimes, the C–C pair correlation function is identical to the hard-template case (Figure S3, right).

Beyond its methodological interest, this soft-template protocol opens the way to applications in which pore geometry must be controlled precisely. In a related work \cite{Waysenson_2026}, we used it to generate carbide-derived-carbon-like electrodes with sub-nanometer pores systematically tuned between $7$ and $11$~\AA, by adjusting the template interaction parameters to target specific pore sizes. This control enabled a systematic investigation of the effect of pore size on supercapacitor performance, with results in quantitative agreement with experiments.

\section{Conclusion}
\label{sec:conclusion}

Starting from a common melt--quench--anneal protocol driven by an MLIP, we have introduced a hierarchy of templating strategies that progressively decouples structural descriptors that are intrinsically correlated in purely thermal protocols. An annealing hold at 3900~K first provides a well-graphitized, low-defect reference carbon. Inserting a rigid cylindrical template that interacts with carbon through a Lennard-Jones potential then provides two separate control parameters. The interaction range $\sigma$ sets the size of the percolating pore, raising the PLD from below 6~\AA\ in the untemplated carbon to above 10~\AA\ for $\sigma = 7$~\AA. The well depth $\varepsilon$ governs the organization of the surrounding walls, which range from ragged, disordered boundaries in the weakly attractive regime to concentric, nanotube-like graphitic shells in the strongly attractive one. Generalizing the template to a spring-connected chain with bending interactions adds tortuosity as a third, independent design variable. The percolating diameter remains essentially set by $\sigma$ regardless of tortuosity, whereas the response of the rest of the pore network depends on $\varepsilon$. Weakly attractive templates promote the opening of larger cavities as tortuosity increases. Strongly attractive templates instead confine the carbon into well-defined shells, which keeps both the LCD and the PSD narrowly constrained. Finally, allowing the template to evolve during the quench introduces a reciprocal coupling between the template and the graphitizing matrix. The carbon network locally reshapes the chain, which sharpens the pore boundaries in the weakly attractive regime and leaves the already well-structured, strongly attractive regime largely unchanged. Throughout, these constraints act on pore geometry while leaving the local structure of the carbon matrix (sp$^2$ fraction, ring statistics and pair correlation shown in Table S1, Figures S1–S3) close to its untemplated values.

The ability to vary one structural descriptor at a time enables controlled numerical experiments on the structure--property relationships of carbon electrodes. Pore size can be tuned systematically, and tortuosity, recently identified experimentally as a key factor governing supercapacitor charging rates \cite{Kress_2026}, can now be addressed in the same way. Combined with constant-potential molecular dynamics \cite{Merlet_2012, Pean_2013, Coretti_2022}, electrodes of prescribed tortuosity at fixed pore size and disorder could rationalize this effect at the molecular scale. Beyond supercapacitors, such models are relevant wherever pore geometry and connectivity govern performance, from capacitive deionization \cite{simoncelli2018a} and CO$_2$ capture \cite{Shi_2023} to sodium storage in hard carbons \cite{Closedpores2025, WangDing2026}.

Methodologically, multiple or branched templates could be combined to build hierarchical pore networks, and the soft-template variant is a first step toward reproducing experimental soft templating with self-assembling block copolymers or surfactants \cite{Liang2004, Meng2005}. More generally, embedding external geometric and interaction constraints into a chemistry-resolving quench should transfer to other disordered porous materials whose structural descriptors are similarly entangled.

\begin{acknowledgement}
Z.W. and B.B--G. have contributed equally to this work. This work was granted access to the HPC resources of IDRIS under the allocation (2026-A0200901387)
made by GENCI. This work was supported by the French National Research Agency under the France 2030
program (Grant No.\ ANR-22-PEBA-0002). B.B-G. acknowledges funding from the french Direction Générale de l'Armement (DGA).
A.M.S. is particularly pleased to contribute to this Festschrift in honor of Mike Klein and remains deeply grateful for his mentorship during his postdoctoral years, many years ago, and for the lasting influence that Mike’s scientific vision, generosity, and example have had on his own path as a scientist.
\end{acknowledgement}
\clearpage

\providecommand{\latin}[1]{#1}
\makeatletter
\providecommand{\doi}
  {\begingroup\let\do\@makeother\dospecials
  \catcode`\{=1 \catcode`\}=2 \doi@aux}
\providecommand{\doi@aux}[1]{\endgroup\texttt{#1}}
\makeatother
\providecommand*\mcitethebibliography{\thebibliography}
\csname @ifundefined\endcsname{endmcitethebibliography}
  {\let\endmcitethebibliography\endthebibliography}{}

\end{document}